# Data-Driven Modeling of Geometry-Adaptive Steady Heat Transfer based on Convolutional Neural Networks: Heat Convection

Jiang-Zhou Peng[1], Xianglei Liu[2], Nadine Aubry[3], Zhihua Chen[1], Wei-Tao Wu[4*]

1. Key Laboratory of Transient Physics, Nanjing University of Science and Technology, Nanjing, 210094, China

2. School of Energy and Power Engineering, Nanjing University of Aeronautics and Astronautics, Nanjing 210016, China

3. Department of Mechanical Engineering, Tufts University, Medford, MA, 02155, USA

4. School of Mechanical Engineering, Nanjing University of Science and Technology, Nanjing, 210094, China

**Abstract**

Heat convection is one of the main mechanisms of heat transfer and it involves both heat conduction and heat transportation by fluid flow; as a result, it usually requires numerical simulation for solving heat convection problems. Although the derivation of governing equations is not difficult, the solution process can be quite complicated and usually requires numerical discretization and iteration of differential equations. In this paper, based on neural networks, we develop a data-driven model for extremely fast prediction of steady-state heat convection of a hot object with arbitrary complex geometry in a two-dimensional space. According to the governing equations, the steady-state heat convection is dominated by convection and thermal diffusion terms, thus the distribution of the physical fields would exhibit stronger correlations between adjacent points. Therefore, the proposed neural network model uses Convolutional Neural Network (CNN) layers as the encoder and Deconvolutional Neural Network (DCNN) layers as the decoder. Compared with fully connected (FC) network model, the CNN based model is good for capturing and reconstructing the spatial relationships of low-rank feature spaces, such as edge intersections, parallelism and symmetry. Furthermore, we applied the signed distance function (SDF) as the network input for representing the problem geometry, which contains more information compared to a binary image. For displaying the strong learning and generalization ability of the proposed network model, the training dataset only contains hot objects with simple geometries: triangles, quadrilaterals, pentagons, hexagons and dodecagons; while the validating cases use arbitrary and complex geometries. According to the study, the trained network model can accurately predict the velocity and temperature field of the problems with complex geometries which has never been seen by the network model during the model training; and the prediction speed is four orders faster than the CFD. The ability of the accurate and extremely faster prediction of the network model suggests the potentials of applying such kind of models to the applications of real-time control, optimization, and design in future.

**Keywords**: Heat transfer, Heat convection, Data-driven model, Convolution neural networks, Signed distance function

(The code will be available upon the publication of the manuscript: https://github.com/njustwulab)

# 1. Introduction

There are many applications of forced convection in daily life and industry. Attributed to the fast development of computational technique and computational ability, numerical simulations have been one of the main methods for solving complex convective heat transfer problems. It is well-known that numerically solving differential equations of heat convective is time consuming, which may become prohibitive for optimization problems involving large number of design parameters. While during early stage of design/optimization it usually does not require high-fidelity simulation results, what favored is that the numerical prediction should be fast for quick iteration. A popular strategy is to use the framework of reduced order modeling (ROM) to enable a fast fluid flow and heat transfer predictions [1]–[3].

In general, a ROM attempts to convert and represent the high dimension dynamic system onto the linear subspace domain by choosing of an appropriate transforming coordinate system [4][5]. An important feature of this transformed space is that it allows for decoupling of spatial and temporal modes effectively [6]. The most common choice for the construction of these spatial transformation bases are the method of proper orthogonal decomposition (POD). The POD method has provided powerful tools building ROM for fluid flow and heat transfer problems [7]–[9]. Indeed, the low computational expense and small memory requirement of this method make it particularly suitable for optimization and aftertreatment analysis. However, the POD method is also limited because it is a linear combination of eigenvectors and eigenvalues and does not explicitly account for the nonlinear interactions of the highly nonlinear dynamic system. That is to say the POD method is significantly effective for quasi-steady-state, time-periodic problems, but might be challenging for highly non-stationary and nonlinear problems [10].

In recent years, the machine learning (ML) and deep learning (DL) based techniques/methods has been showing power on building surrogate/reduced-order model for highly nonlinear dynamic problems, including applications of aerodynamics, heat transfer and fluid flow. The big difference between DL enabled and the POD based ROM is that DL can be used to directly setup a nonlinear relationship between abundant of inputs and outputs of a target system. This process of fitting to available data, known as model training, generates a low-dimensional subspace which records the mean behavior of the underlying phenomena of flows; and this training process allows for the representation of complex relationships/features which cannot be expressed explicitly in a functional form [11]. In practice, the DL enabled ROM has been demonstrated to be able to accurately capture the spatial and temporal nonlinear features of fluid flow. Wang et al. [12] presented a model identification of reduced order fluid dynamic systems by using deep learning, and proved the framework is capable of capturing the features of the complex fluid dynamics with less computational cost. Fukami et al. [13] used machine learning to perform super-resolution analysis of grossly under-resolved turbulent flow data and reconstructed the high-resolution flow field; that is their model successfully builds a nonlinear mapping between low-resolution and high-resolution turbulent field. Alternatively, an approach known as DL based closure modeling has been used to improve the performance of the traditional ROM methods [14]–[16].

The deep learning (DL) based reduced order modeling method has started to cause the attention of thermal engineering community, although there are only a few of publications available. San and Maulik [6] applied machine learning method for developing a data-driven closure modeling for stabilizing projection based reduced-order models for the Bousinessq equations; that is improving the performance of the traditional ROM by DL. Gao et al. [18] proposed a physics-constrained CNN architecture to learn solutions of parametric PDEs on irregular domains, where the PDEs include heat transfer equation and Navier-Stokes equations. Their results demonstrate the effectiveness of the proposed DL approach in predicting the temperature field and velocity field.

In current study, we apply CNN to build a reduced-order, geometry-adaptive and steady-state heat convection model, since CNN has demonstrated strong feasibility in geometry representation and per-pixel prediction in two-dimensional fields [19]–[21]. In section 2, we introduce the network architecture of the reduce order model, the preparation of the datasets and the training algorithm. In section 3 and 4, we present and discuss outstanding performance of the network model.

## 2. Methods

In this paper, we propose a CNNs based ROM directly builds a mapping between physical fields and signed distance function (SDF) which represents the problem geometry. The networks and its training are implemented using Tensorflow. The "ground truth" (training dataset) is generated through CFD simulation using OpenFOAM.

### 2.1. Design of CNNs-ROM framework

#### 2.1.1. Framework workflow

Fig. 1 depicts the mainly graph of the proposed reduced order model, which shows the learning/training strategy and the method of a new prediction using the network model. For the model training, we use signed distance function (SDF) which carries physical information of the geometry as the network input, and the results (temperature/velocity) of numerical simulations as the learning target (output/label) of the CNNs. After training with proper amount of dataset, the trained CNNs can predict the temperature/velocity fields based on the new SDF values.

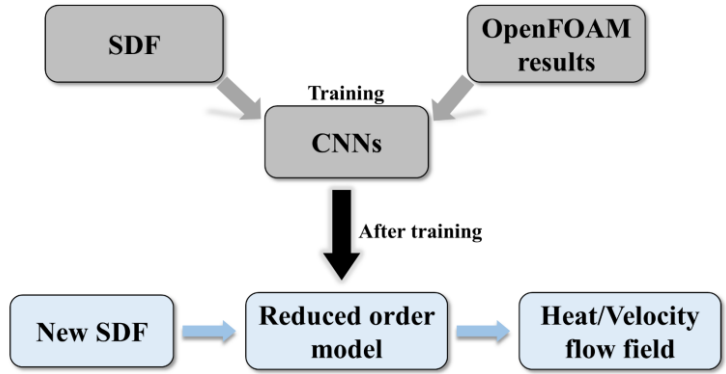

Fig. 1 Schematic of CNNs-ROM framework.

#### 2.1.2. Signed distance function

The signed distance function (SDF) is proposed as the geometry representation at the network input. Using the SDF, the geometry of the object is represented as a level set function defined over the simulated space in which the object is embedded [22], [23]. Compared to the boundaries and geometric parameters representation, the SDF provides more physical and mathematical information for CNNs. In SDF representation method, the zero-level set is created to represent the location of the boundary of the object. That is, the boundary of the object, $\sum$, is represented as the zero-level set of a continuous level set function '$\phi$' defined in a domain $\Omega \subset R^2$; i.e.,

$$\Sigma = \{X \in R^2 : \phi(X) = 0\} \tag{1}$$

The level set function $\phi(X)$ is defined everywhere in the domain $\Omega$, and $\phi(X) = 0$ if and only if $X(x_i, y_i)$ is on the object boundary, such as the edge of the polygon shown in Fig. 2. The SDF associated to a level set function $\phi(X)$ is defined as,

$$D(X) = \min_{X_2 \in \Sigma} |X(x_i, y_i) - X_2(x', y')| sign(\phi(X)) \tag{2}$$

$D(X)$ is an oriented distance function, and the sign function "$sign$" is defined as:

$$sign(x) = \begin{cases} 1 & if\ x > 0 \\ 0 & if\ x = 0 \\ -1 & if\ x < 0 \end{cases} \tag{3}$$

By using the above level-set function, one can represent an arbitrary shape in the fixed design domain, $\Omega$ [24]. Fig. 2 gives the distribution of the SDF representation of a training dataset.

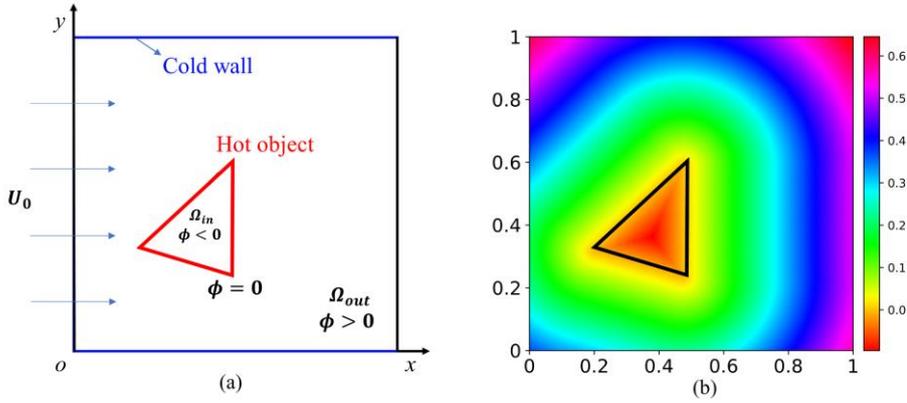

Fig. 2 (a) Schematic of the object and the studied domain; (b) Distribution of the SDF representation of a triangle, where the boundary of the triangle is in black, and the magnitude of the SDF values equals to the minimal distance of the space point to the triangle boundary.

### 2.1.3. Architecture of CNNs

The proposed network architecture consists of convolution layers and transpose convolution (also called deconvolution) layers. To ensure the high nonlinearity of the proposed reduced-order model, we use a multilayer deep structure for our neural network [25]. In the model, multiple convolutional layers are applied to extract a highly encoded [26] geometry representation from the SDF (input matrix of the network), and the encoded geometry representation is decoded by multiple deconvolutional layers [27] to predict the physical fields. The choice of the network structure is highly dependent on the problem, the data quality and even the dataset size [28]. For example, it would be necessary to build deep convolutional neural networks in our problem due to the big size and complexity of the training dataset. Fig. 3 shows the structure and components of the CNNs model, and Table 1 displays the parameters of each layer of the network model. We will describe each part of the network in the following subsection.

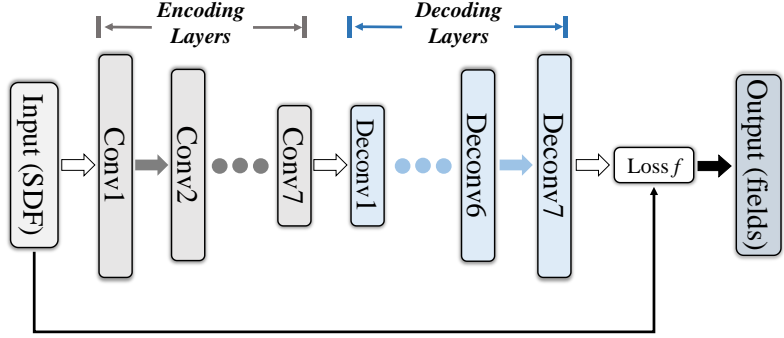

Fig. 3 Architecture of the CNNs-based reduced-order model. 'Conv' denotes the convolutional layer; 'Deconv' denotes the deconvolution layer.

Table 1 Parameters of the neural network model of each layer, where $w \times h$ denotes the size of convolution kernel ($F$); $n_F$ denotes the number of $F$; $N_l \times N_l \times n_F$ is the size of the $l$th layer after operation.

| Name of Layer | Executing operation $w \times h \times n_F/\text{stride}$ | Shape $N_l \times N_l \times n_F$ |
|---|---|---|
| **Input of model** | -- | 250×250×1 |
| **Conv1** | 2×2×32/2 | 125×125×32 |
| **Conv2** | 2×2×64/2 | 62×62×64 |
| **Conv3** | 2×2×128/2 | 31×31×128 |
| **Conv4** | 2×2×256/2 | 15×15×256 |
| **Conv5** | 2×2×512/2 | 7×7×512 |
| **Conv6** | 2×2×512/2 | 3×3×512 |
| **Conv7** | 2×2×1024/1 | 2×2×1024 |
| **Deonv1** | 2×2×1024/1 | 3×3×1024 |
| **Deonv2** | 3×3×512/2 | 7×7×512 |
| **Deonv3** | 3×3×256/2 | 15×15×256 |
| **Deconv4** | 3×3×128/2 | 31×31×128 |
| **Deconv5** | 2×2×64/2 | 62×62×64 |
| **Deconv6** | 3×3×32/2 | 125×125×32 |
| **Deconv7** | 2×2×1/2 | 250×250×1 |
| **Output** | Reshape | 250×250 |

### 2.1.4. Encoding part

The encoding part aims to reducing the dimension of input data and extracting the potential spatial features between neighboring points of SDF matrix and convolution layers. According to the governing equations, the steady-state heat convection is dominated by convection and thermal diffusion terms, thus the distribution of the physical fields would exhibit stronger correlations between adjacent points. As the CNNs are very powerful for handing two dimensional data with locality structure [29], [30], in current approach the data-driven model is developed using the CNNs. Fig. 4 shows the schematic of the convolution (feature extraction) operation by the network.

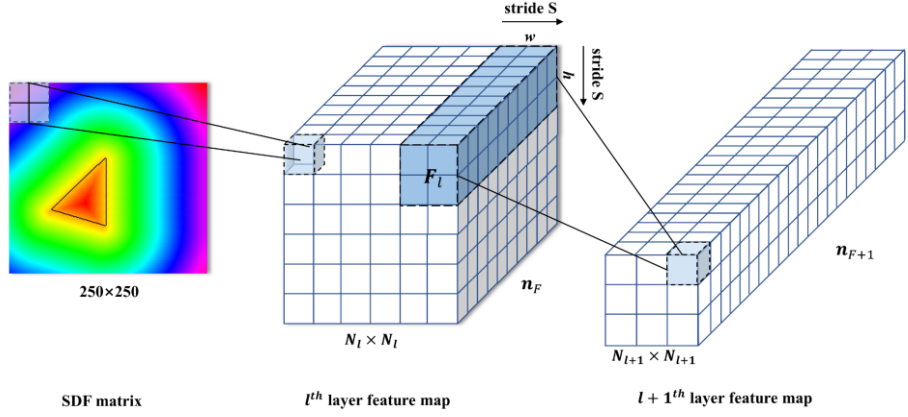

Fig. 4 Schematic of the convolution operation. The light blue matrix denotes a 2×2 convolutional kernel, and the white matrix represents the feature map.

In Fig. 4, the $2 \times 2 \times n_F$ light blue matrix denotes the convolutional kernel, the white matrix is the feature map matrix (also called next layer's input matrix) and the blue translucent matrix represents the output of the convolution operation. Besides the size of the kernel, $F(w \times h)$, the convolutional output is also influenced by the kernel stride, $S$, the size of the channels, $n_F$, and the padding size, $P$. The padding operation adds zeros around the border of the input matrix, and the kernel stride controls the sliding step size of the kernel. The size of the output matrix after the convolutional operation can be calculated as,

$$N_{l+1} = \frac{N_l - F + 2P}{S} + 1 \tag{4}$$

where $H_l$ is the size of feature map at $l^{th}$ layer. In current work, zero padding size ($P$=0) is used. From the above equation, it can be seen that after several convolutional operations, the size of the original input can be reduced significantly, and the features of the original input is also highly encoded, thus the memory space required by CNNs training was reduced.

To increase the non-linear capability of CNNs each convolution or deconvolution layer involves the nonlinear activation operations. Mathematically the nonlinear activation operation can be expressed as,

$$\boldsymbol{a}_l = \sigma(\boldsymbol{W}_l * \boldsymbol{a}_{l-1} + \boldsymbol{b}_l) \tag{5}$$

where $\boldsymbol{W}_l$ is the weights or convolutional kernel of the current layer, $\sigma$ denotes the nonlinear activation function, $\boldsymbol{a}_l$ and $\boldsymbol{a}_{l-1}$ are the input and the output of the layer respectively, $*$ is the convolutional operator, and $\boldsymbol{b}_l$ is the bias term. The introduction of $\sigma$ is crucial for neural networks to possess non-linearity. In this paper, we apply the Rectified linear unit (RELU) activation function,

$$\sigma(x) = \begin{cases} 0, x < 0 \\ x, x \geq 0 \end{cases} \tag{6}$$

The advantages of RELU are that its computation cost is cheap as the function has no complicated math, and it converges fast thus the model takes less time to train or run. It should be noticed that, because the final prediction of the network model needs to be a continuous regression, there is no activation function between the output layer and the 'Deconv7' layer.

### 2.1.5. Decoding part

The decoding part is designed to analyze the highly encoded features and decode them to be a recognizable velocity and temperature fields. The decoding part is composed by multiple deconvolutional layers, and these inverse operations (deconvolution) unravel the high-level features encoded by the encoding part. Fig. 5 shows the schematic of the deconvolution operation of one channel.

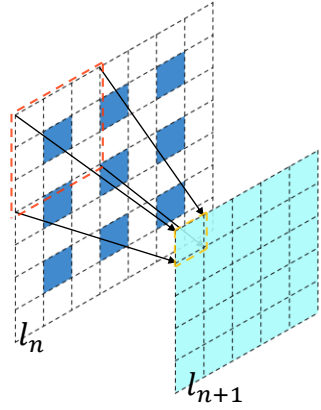

Fig. 5 Schematic of the deconvolutional layer: the dark blue block represents the feature map at the $l_n$ layer, the white block illustrates padding location of $l_n$ layer's feature map, the pool blue block represents the feature map at next layer ($l_{n+1}$).

The deconvolution operation needs appropriate basis functions to learn invariant subspaces of feature map, and then compensates the minutia information to predict the physical fields by multilayer iterative optimization. For instance, as shown in Fig. 5, for adjusting the width and height of the output feature map (pool blue block), random values (optimized by the algorithm) will be padded at the white block around the block (dark blue block) of the input feature map. Therefore, highly encoded information is recovered by applying the deconvolutional decoder. Meanwhile, the CNNs learns/trains its ability of predicting physical fields based on the SDF through this process.

**2.2. Model training**

The model training is an iterative process, which continuously minimizes the loss between the outputs of the network predicted ($\hat{\psi}$) and the ground truth ($\psi$), for obtaining the optimal model parameters, $\boldsymbol{\theta}$. Further, we need to condition our ROM prediction based on the SDF, as the prediction inner the hot object is unphysical and should not affect the loss function. The SDF based conditioning operation is defined as follow,

$$J = \frac{1}{N}\sum_{n=1}^{N}\left(\left(\psi_n(x,y) - \hat{\psi}_n(x,y)\right) \cdot \delta(x,y)\right)^2 + \lambda\|W\|_2 \quad (7)$$

$$\delta(x,y) = \begin{cases} 1, \phi(x,y) \geq 0 \\ 0, \phi(x,y) < 0 \end{cases} \quad (8)$$

where $(x, y)$ is the index of the space point and $n$ is the index of the case number, $N$ is the size of the (batch) dataset, $\psi$ indicates the result by the numerical simulation, $\hat{\psi}$ is the result predicted by the network model, $W$ denotes all the weight of the network layers, $\lambda$ is the regularization coefficient, and $\lambda\|W\|_2$ is the L2 Regularization term for preventing model overfitting. Such a conditioning operation eases the training process and improve the prediction accuracy [31].

To train the model efficiently, a minibatch-based learning strategy is used. Small batch training has been shown to provide improved generalization performance and reduces the memory cost significantly [32]. The loss function is computed using a randomly selected subset (i.e. the batch size *n*) and the whole optimization process forms an iteration way. Moreover, the Adaptive Moment Estimation (Adam) method is implemented as the optimization algorithm for the model training; Adam method has been proved to be robust and well-suited for large datasets and/or optimization problems of high-dimensional parameter spaces [33]. The hyper-parameters of the optimization algorithm for the network training are shown in Table 2.

Table 2 Hyper-parameters of the optimization algorithm.

| Hyper-parameter | Value |
|---|---|
| Batch size | 64 |
| $\lambda$ | 0.0001 |
| $\alpha$ | 0.00008 |

**2.3. Preparation of dataset**

The training and test datasets consist of 5 types of 2D simple hot objects in a square simulated domain (see Fig. 2 as an example): triangles, quadrilaterals, pentagons, hexagons and dodecagons. With the simulated domain kept unchanged, each set of the hot objects are randomly different in size, shape, orientation and location. In general, it is more often that the object locates near the center of the studied domain, therefore the location of the object is generated following a normal distribution in the function of *x* and *y* coordinates and using center of the domain as the mean. Each 2D simple object is preprocessed into a 250×250 pixel matrix based on the SDF. The training and testing dataset totally contain 10,000 samples (2,000 random samples for each type of object), where the testing dataset contains 1,000 samples (random separated from the whole training and testing dataset). The simulated domain is discretized using an unstructured mesh tool, SnappyHexMesh [34], [35], and the numerical simulation is performed using OpenFOAM, where SIMPLE algorithm is used. Regarding validation dataset, which is dedicated to validate and indicate the generalization ability of our CNNs model, more complex geometries of the hot objects are chosen, see Fig. 6. Those samples have never been seen by the network model during the training process. All the process of the mesh generation and numerical simulation for generating the validation dataset are same to the training dataset.

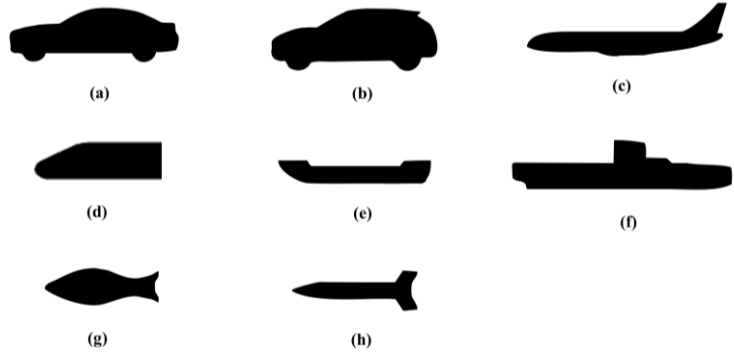

Fig. 6 Geometry of the objects of the validation dataset, (a) Car, (b) Car2, (c) Airplane, (d) Locomotive, (e) Ship, (f) Submarine, (g) Bionic Fish, (h) Missile.

The physics or governing equations of the steady-state heat convection can be given as,

$$\nabla \cdot \boldsymbol{v} = 0 \tag{9}$$

$$(\boldsymbol{v} \cdot \nabla)\boldsymbol{v} = -\nabla p + \nu \nabla \cdot \nabla(\boldsymbol{v}) \tag{10}$$

$$(\boldsymbol{v} \cdot \nabla)T = \alpha \nabla \cdot \nabla(T) \tag{11}$$

where $\boldsymbol{v}$ is the velocity vector, $p$ is the pressure, $T$ is the temperature, $\nabla \cdot$ is the divergence operator, and $\nabla$ is the gradient operator. $\nu$ and $\alpha$ are the momentum and thermal diffusivity. It should be noticed that the viscous dissipation has been ignored. Based on the following parameters:

$$\boldsymbol{v}^* = \frac{\boldsymbol{v}}{v_0}; \; T^* = \frac{T - T_0}{T_1 - T_0}; \; P = \frac{p}{v_0^2}; \; Re = \frac{v_0 L_r}{\nu}; \; Le = \frac{\alpha}{v_0 L_r}; \; \nabla^* \cdot \coloneqq L_r \nabla \cdot; \nabla^* = L_r \nabla$$

The governing equations can be normalized as follows:

$$\nabla \cdot \boldsymbol{v} = 0 \tag{12}$$

$$(\boldsymbol{v} \cdot \nabla)\boldsymbol{v} = -\nabla P + \frac{1}{Re} \nabla \cdot \nabla(\boldsymbol{v}) \tag{13}$$

$$(\boldsymbol{v} \cdot \nabla)T = Le \nabla \cdot \nabla(T) \tag{14}$$

where $L_r$ is a reference length, $T_0$ is the temperature of the cold wall and $T_1$ is the temperature of the hot object, $v_0$ is the inlet mean velocity, $Re$ is the Reynolds number and $Le$ is the Lewis number which is the ratio of thermal diffusivity to convective mass transport. Notice that the asterisks have be ignored for simplicity. In the following study, we keep the Reynolds number and the Lewis number as 10 and 15, respectively.

## 3. Results

We use the average, mean and maximum relative error to measure the prediction accuracy of the network model. For a studied case, the relative error for velocity and temperature is calculated as,

$$E(x,y) = \frac{|v(x,y) - \hat{v}(x,y)|\delta(x,y)}{v_0}; \; E(x,y) = \frac{|T(x,y) - \hat{T}(x,y)|\delta(x,y)}{|T(x,y)|} \tag{15}$$

The maximum point relative error is defined as,

$$E_{max} = \max(E(x,y)) \tag{16}$$

The mean relative error is defined as,

$$E_{mean} = \frac{\sum_x \sum_y E(x,y)}{\sum_x \sum_y \delta(x,y)} \tag{17}$$

For evaluating the performance of the network model over several different cases, we also define average relative error as,

$$E_{avg} = \frac{1}{N} \sum_{n=1}^{N} (E_{mean})_n \tag{18}$$

where $n$ is the index of the studied cases. Furthermore, from equation (15), it should be noticed that when calculating the error, we only consider the domain outside the hot object.

### 3.1. Feasibility of the CNNs based model

In this section, we verify the feasibility of the proposed framework from two aspects: 1) the outstanding performance of the CNNs model on predicting the velocity and temperature fields; 2) the advantage of SDF representation.

### 3.1.1. Performance of the temperature field prediction

We first evaluate the prediction performance of the temperature field of the network model. After proper model training, the prediction accuracy of the testing dataset approaches to be higher than

98.75%. Fig. 7 visualizes two typical temperature fields of the test dataset predicted by the network model and numerical simulation (the OpenFOAM), and the corresponding relative error distribution, see Fig. 7. We can see that the network model has given a satisfied accurate prediction on temperature field. Furthermore, the figure of the relative error distribution indicates that the large error mainly locates in the region with large temperature gradient.

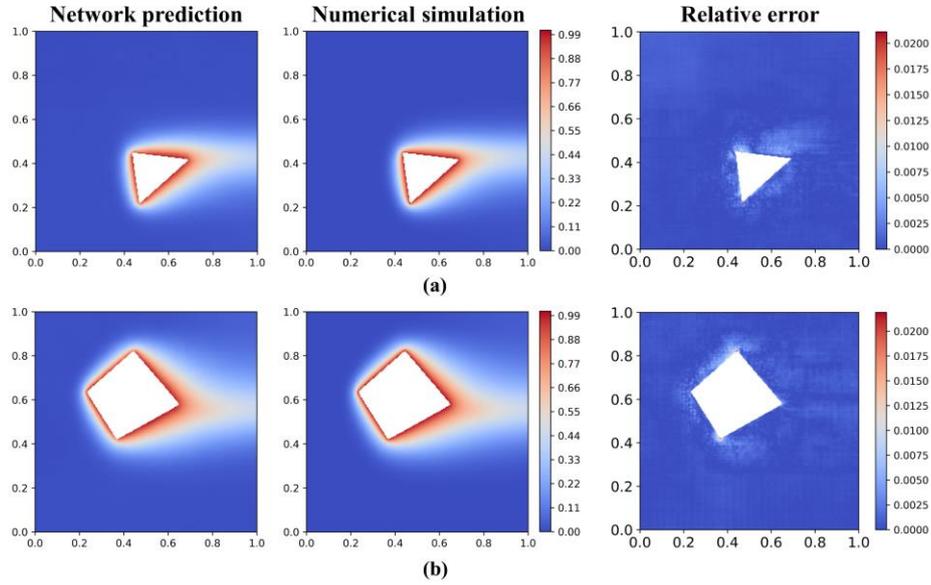

Fig. 7 Temperature field and error distribution of validation cases. (a) Triangle, and (b) Quadrangle. The first and second column shows the temperature field predicted by the network model and the numerical simulation (OpenFOAM). The third column shows the relative error distribution.

Fig. 8 shows the temperature fields of the validation cases by the network model and the numerical simulation. The displayed validation cases include: Car, Car2, airplane and Locomotive. For the cases shown in Fig. 8, the maximum relative error ($E_{max}$) is lower than 5.2%; the large error happens more frequently near the hot objects, especially at the boundary with large curvature. Fig. 9 shows temperature profiles distribution along *y*-direction at different x-position, by the network model (symbols) and numerical simulation (lines). The good coincident of the symbols and lines quantitatively affirms the accuracy of the network model. From the results, the comparatively poor prediction performance appears in the wake (the x=0.65, blue line), but the error is still within the acceptable range.

Overall, considering that the network model has only "seen" triangles, quadrilaterals, pentagons, hexagons and dodecagons during the training process, such a high prediction accuracy on the validation cases with much complex geometries indicates that the network model has an outstanding ability of the generalization and extensionality.

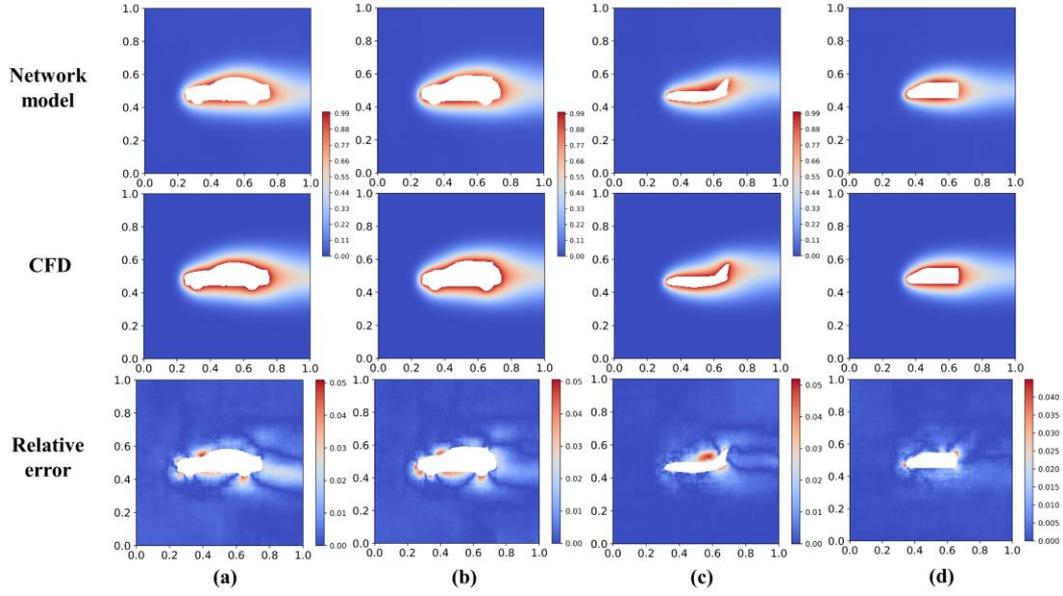

Fig. 8 Temperature field and relative error distribution of the validation cases with geometry of (a) Car, (b) Car2, (c) Airplane, (d) Locomotive.

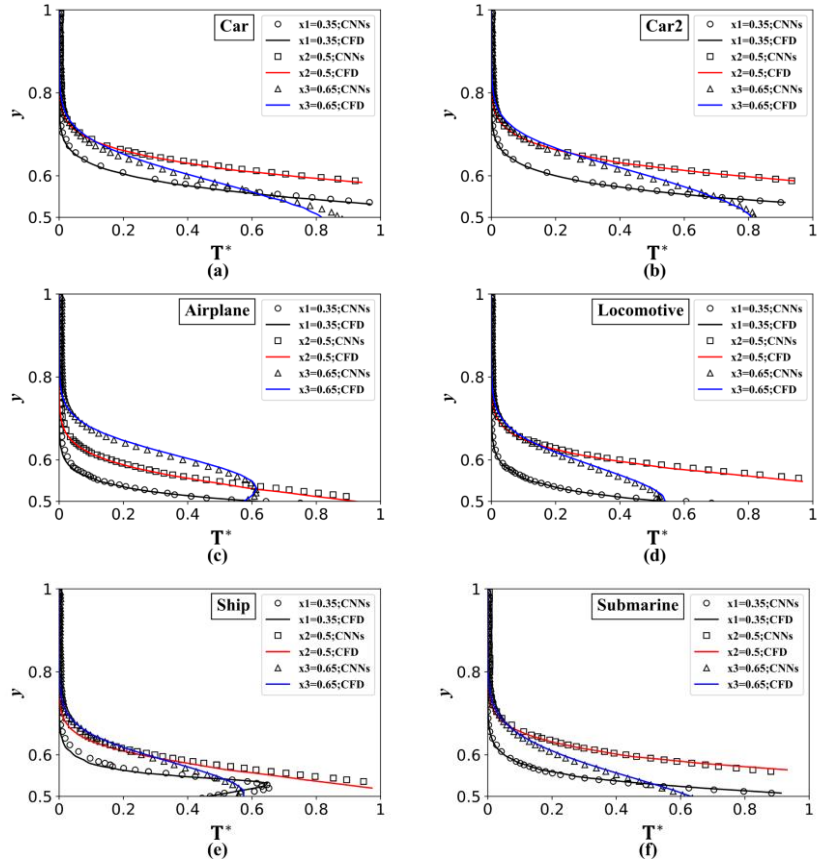

Fig. 9 Temperature profiles of the validation cases distribution along *y*-direction at $x_1=0.35$; $x_2=0.5$; $x_3=0.65$, by the network model (symbols) and numerical simulation (lines). (a) Car, (b) Car2, (c) Airplane, (d) Locomotive, (e) Ship, (f) Submarine.

Furthermore, we study the computational cost of the field prediction by the network model and CFD. Table 3 quantitatively compares the time consumption for predicting the steady-state temperature field by the network model and the numerical simulation. Overall, the network model can speed up the prediction for 4 orders; and as the geometry of the problem becomes more complex, that is it requires finer mesh for CFD to converge the simulation, the prediction speedup by the network model becomes more outstanding. In addition, we also compare the prediction time cost by the network model using different GPUs, 2080ti and 1660s, see Table 4. From the table, we can see that the performance of these two equipment has no big difference.

Table 3 Time consumption for predicting the steady-state temperature field by the network model (GPU, 2080ti) and the numerical simulation (CPU, OpenFOAM).

| Geometry object | CNNs (s) | OpenFOAM (s) | Grid quantity | Speedup |
|---|---|---|---|---|
| Car | 0.235399 | 5054 | 10144 | $2.15 \times 10^4$ |
| Car2 | 0.242347 | 4714 | 10000 | $1.95 \times 10^4$ |
| Airplane | 0.231412 | 3506 | 9032 | $1.52 \times 10^4$ |
| Locomotive | 0.228387 | 2193 | 7758 | $0.96 \times 10^4$ |
| Bionic Fish | 0.24634 | 3530 | 8252 | $1.43 \times 10^4$ |
| Missile | 0.232379 | 3641 | 8528 | $1.57 \times 10^4$ |
| Ship | 0.236368 | 3516 | 8956 | $1.49 \times 10^4$ |
| Submarine | 0.23936 | 3993 | 9876 | $1.67 \times 10^4$ |

Table 4 Time consumption for predicting the steady-state temperature field by the network model on GPU 1660s and GPU 2080ti.

| Geometry object | 1660s time (s) | 2080ti time (s) |
|---|---|---|
| Car | 0.398289 | 0.235399 |
| Car2 | 0.3723 | 0.242347 |
| Airplane | 0.352322 | 0.231412 |
| Locomotive | 0.379252 | 0.228387 |
| Bionic Fish | 0.319414 | 0.24634 |
| Missile | 0.316422 | 0.232379 |
| Ship | 0.31742 | 0.236368 |
| Submarine | 0.318419 | 0.23936 |

**3.1.2. Performance of the velocity field prediction**

The steady-state velocity fields of the validation cases predicted by the network model and numerical simulation are shown in Fig. 10. From the results, it can be observed that the velocity field by the network model shows good agreement with the CFD simulation. The right column of the Fig. 10 shows the distributions of the relative error: we can see that the maximum error happens frequently in the edge of objects, because the flow field there is usually accompanied by large velocity gradient.

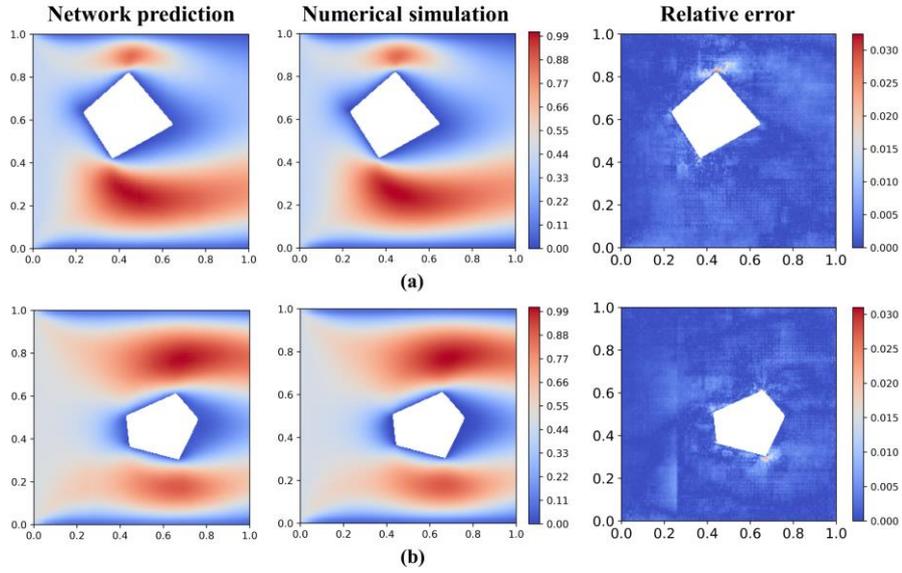

Fig. 10 Velocity field and error distribution of the testing cases. (a) Quadrangle, and (b) Pentagon. The first and second column are the fields predicted by the network model and the numerical simulation (OpenFOAM), respectively. The third column shows the relative error distribution.

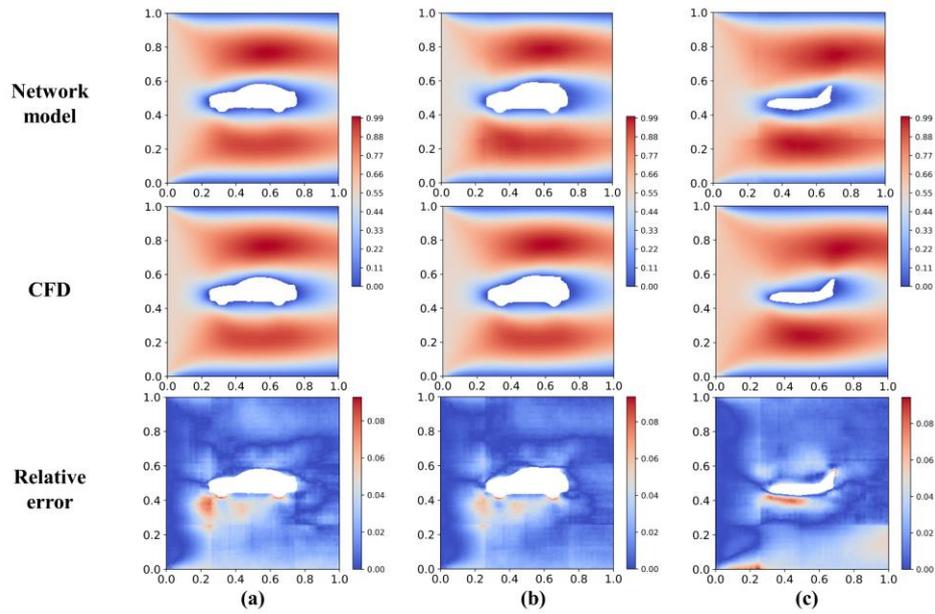

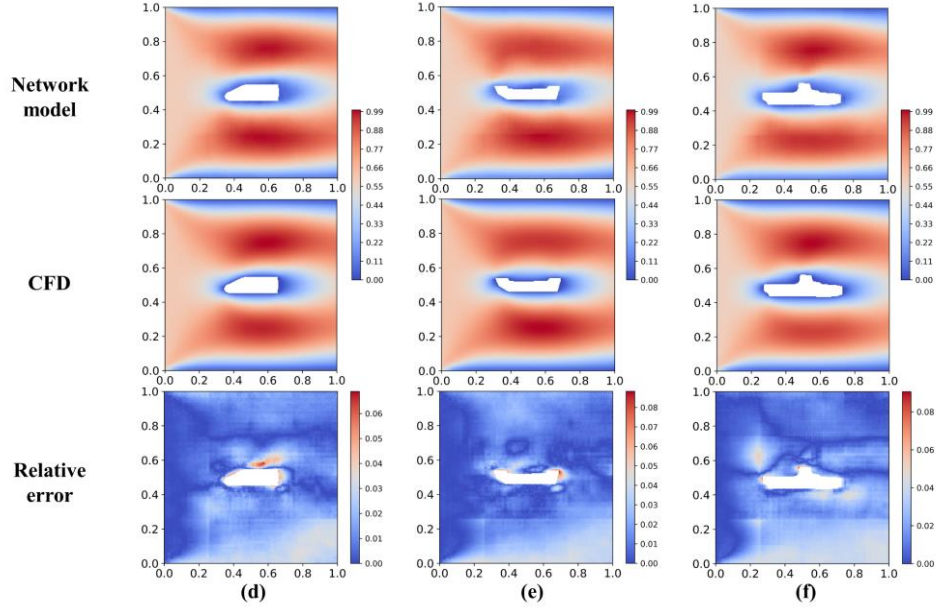

Fig. 11 Velocity field and relative error distribution of the validation cases with geometries of (a) Car, (b) Car2, (c) Airplane, (d) Locomotive, (e) Ship, (f) Submarine.

We further verify the ability of the generalization and extensionality of the network model with validation datasets, see Fig. 6. The velocity fields of the validation dataset predicted by the network model and numerical simulation (OpenFOAM), and the corresponding relative error distribution is shown in Fig. 11. For the six cases shown in the figure, the maximum mean relative error ($E_{mean}$) is less than 2% and the maximum point relative error is less than 10%. In Fig. 12, we quantitatively compared the max ($E_{max}$) and mean relative error of the studied validation cases. Compared to the testing cases, since the complexity of the geometric objects increases, the prediction error of the network model become larger. Furthermore, as the predicted shape gradually deviates from the polygon (Ship, Submarine, e.g.), the prediction error increases. Overall, the low mean error suggests the network model is able to provide accurate prediction on cases with arbitrary complex geometries.

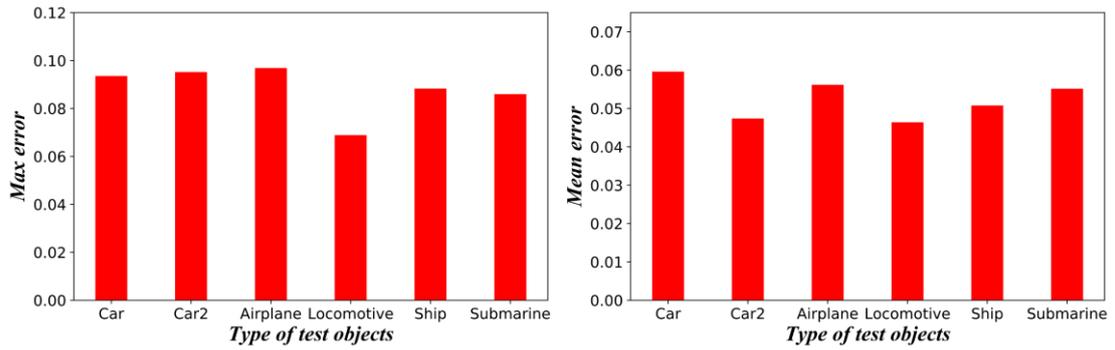

Fig. 12 Max ($E_{max}$) and mean ($E_{mean}$) relative prediction error on the velocity field by the network model for various validation objects.

### 3.1.3. Advantages of the SDF representation

In previous subsections, we have shown the outstanding performance of the proposed network model, where the SDF is used as the geometry representation and the neural network inputs; while according to the literature review, in most previous works a binary images is used as the input of the CNNs [36], [37]. The binary image represents the geometry as follows: its element is 0 if and only if the position is on the object boundary or inside the object. To indicate the effectiveness of the SDF, we trained a model using binary image as the geometric representation with the exact same network architecture. The average relative error ($E_{avg}$) by two different network models on the testing and validation dataset is summarized in the Table 5.

Table 5 Average relative error ($E_{avg}$) on prediction the testing and validation dataset by the network models using the SDF and binary image representation.

| Datasets | Prediction field | SDF | Binary |
| --- | --- | --- | --- |
| Testing | Velocity | 2.79% | 6.56% |
| | Temperature | 0.83% | 4.62% |
| Validation | Velocity | 5.02% | 18.27% |
| | Temperature | 1.91% | 10.82% |

Table 5 indicates that the SDF representation is much more effective than the binary representation. Regarding the validation dataset, the error of using the SDF representation is significantly smaller than the model using the binary representation. This can be attributed to the reason that each value in the SDF matrix carries a certain level of global information, while for a binary image only the values on the boundary of the object carry useful information. Therefore, the SDF representation achieves a much better performance than the binary representation. In the following investigation, only the SDF representation is used as the model input.

**3.2. Influence of the space distribution of the hot object**

In this part, firstly, we investigate the effect of the space distribution of the hot object on the performance of the network model. Although, the network model has shown good prediction accuracy on validation dataset, the predicted objects above are fixed at the center of the entire field. As mentioned in the section of data preparation, the location of the object is generated following a normal distribution in the function of x and y coordinates and using center of the domain as the mean. Hence, in this section, we study the effect of the distance between the center of the hot object and the center of the studied domain. Fig. 13 shows the prediction results by the network model and numerical simulation (The OpenFOAM), and the corresponding relative error. The result shows that the maximum error is less than 8%, which illustrates that the network model is robust to the randomness of the space distribution of the hot object in the entire flow field.

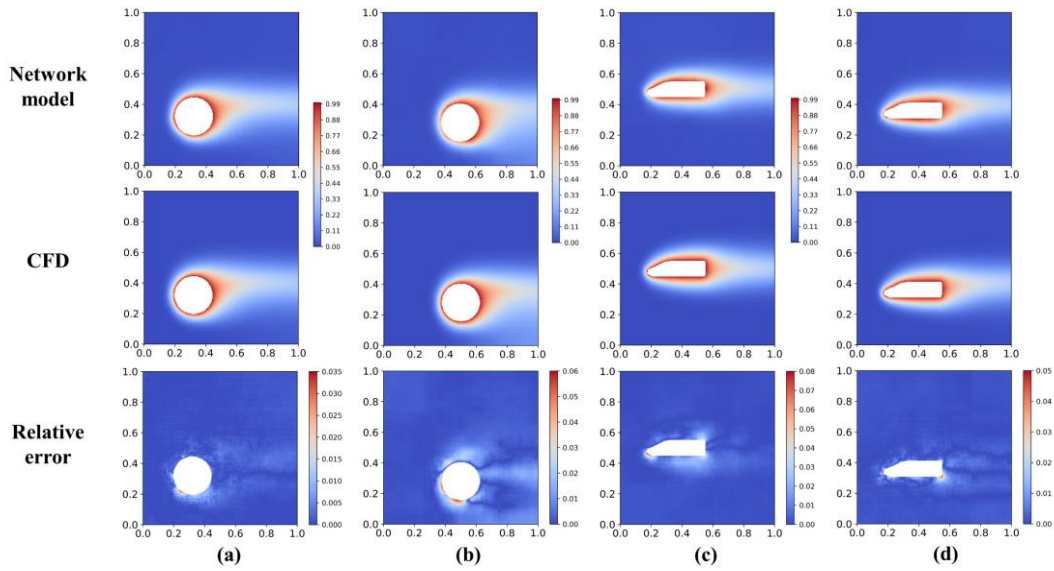

Fig. 13 Temperature distribution predicted by the network model and numerical simulation with different positions of the hot objects, and the corresponding relative error. (a) Left circle, (b) Down circle, (c) Left locomotive, (d) Lower left locomotive.

Secondly, since all the training cases only contain single hot object, here we also investigate performance of the network model on predicting the problems with multiple hot objects. Compared to studying the spatial distribution of the hot object, the problem with multiple hot objects provide a huger challenge to the network model. Fig. 14 shows temperature fields of the case with two circles predicted by the network model and numerical simulation, and the corresponding relative error. It can be observed that the network model can still give a less accurate but still reasonable prediction. Using more training data set to train the network model is an effective solution to further improve the prediction accuracy of the cases with multiple hot objects.

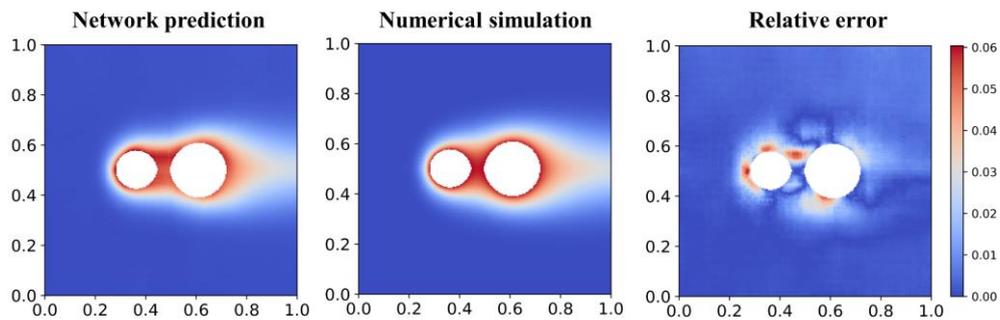

Fig. 14 Temperature fields of the case with two circles predicted by the network model and numerical simulation and the corresponding relative error.

### 3.3. Influence of incorporating velocity into input matrices

For most thermal applications, people are more interested in the temperature than the velocity field. In this section, we study the effect of additionally incorporating velocity field as portion of the input matrices. The prepared cases have the same hyper-parameters and network architecture. Fig. 15 shows the mean ($E_{mean}$) and maximum ($E_{max}$) relative prediction error on the temperature field by the network models with and without incorporating velocity field as the portion of the input matrixes.

Fig. 16 plots the relative prediction error of the two studied situations along y-direction at different x positions. The relative error ($E(x,y)$) is calculated by equation (15). From Fig. 15 and Fig. 16, it can be observed that the overall prediction accuracy increases as the network model further incorporates the velocity information, while the improvement is moderate. Therefore, for current problems, incorporating velocity information is not economic.

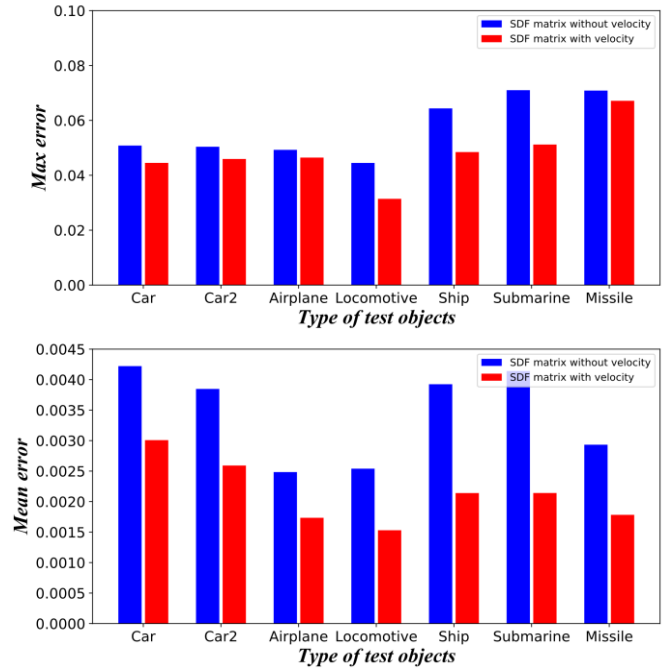

Fig. 15 Max ($E_{max}$) and mean ($E_{mean}$) relative prediction error on the temperature field by the network models with and without incorporating velocity field as the part of the input matrixes. The blue bar denotes the cases without velocity information, and the red bar denotes the cases with velocity information.

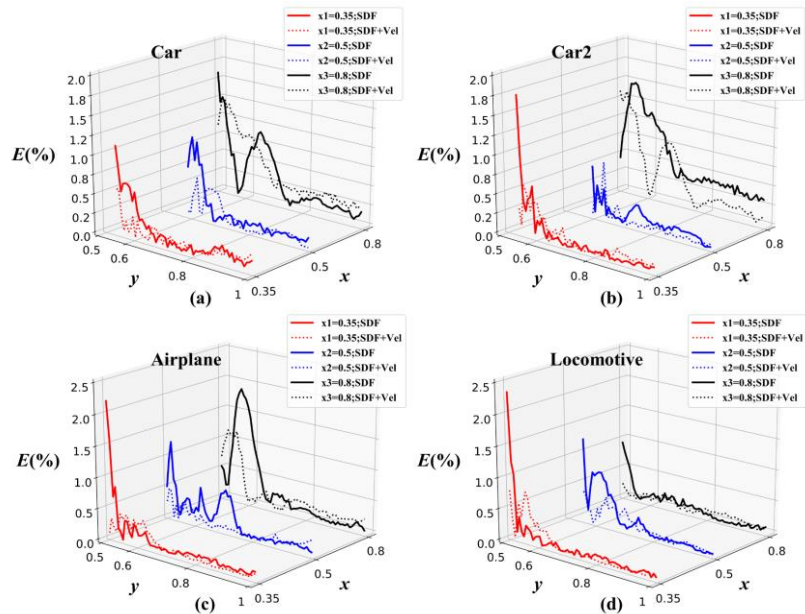

Fig. 16 Relative error profile along y-direction at $x_1$=0.35; $x_2$=0.5; $x_3$=0.8. The studied validation cases are (a) Car, (b) Car2, (c) Airplane, (d) Locomotive.

It can be concluded from above results, additionally incorporating velocity field as portion of the input matrices can enhance the accuracy of the temperature prediction of the neural network. Here, we further investigate the impact of the type of the embedded velocity field on the performance of the neural network model. In the above portion, the input matrix of SDF combined velocity magnitude has been studied, following, we investigate another two types of input matrix: SDF combined x-velocity, SDF combined y-velocity. Fig. 17 shows the training history of the model with different input matrices. Fig. 18 shows the temperature profiles distribution along *y*-direction at the dimensionless x-position of 0.8, predicted by the network model with different input matrices and numerical simulation. It can be seen from Fig. 17 and Fig. 18 that the case of SDF combined velocity shows the fastest convergence speed and the highest accuracy, as it contains most physical information.

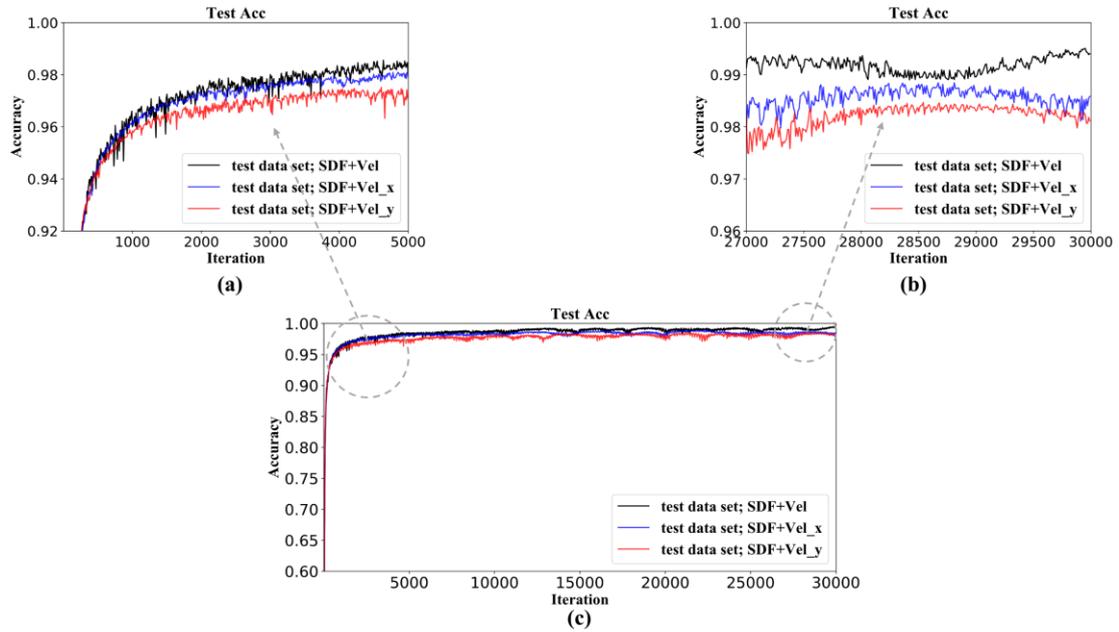

Fig. 17 Training/convergence history of the network models. (a) Local magnification in iteration from 0 to 5000, (b) Local magnification in iteration from 27000 to 30000.

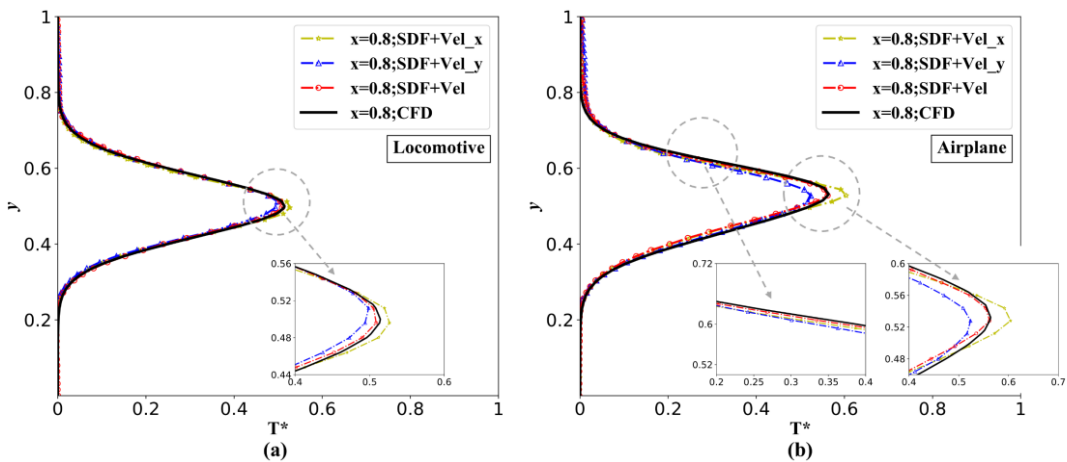

Fig. 18 Temperature profiles of the validation cases distribution along *y*-direction at *x*=0.8, by the network model with different input matrices and numerical simulation. (a) Locomotive, (b) Airplane.

## 4. Discussions

Reduced-order modelling is a data-driven method, and it provides a way of fast predicting physical fields without solving partial differential equations (PDEs) iteratively and numerically. An efficient reduced-order prediction model is useful for the production design and optimization, and the real-time simulation/prediction in some control scenario. The results discussed in last section indicate that the proposed CNNs based reduced-order model achieves high accuracy and robustness, even the geometry of the problem is much more complex than the training dataset. The strong ability of the generalization and extensionality is one of the major advantages of the CNNs. Furthermore, compared to the CFD, the time consumption for the field prediction by the network model can be negligible.

Although the proposed model has shown outstanding extensionality, the prediction accuracy is still affected obviously as the physical conditions (geometric in current paper) gradually deviates from the training dataset; and of course this also applies for traditional data-driven method, such as POD and DMD [38], [39]. Certainly, for relieving this problem, a CNNs-based ROM should be trained using as exhaustive dataset as possible; while it is time-consuming to retrain the network model once we get new/additional data. One solution to this problem is to make full use of the trained reduced-order model for fast reconstruction/retraining of the model.

This is a preliminary attempt to develop the reduced-order model predicting the heat convection fields with arbitrary geometries based on the CNNs. As the beginning of exploration, the studied problem is kept as two-dimensional and laminar, and the performance of the network model is satiated and exciting; while the problems of engineering optimization and design in real applications are usually three-dimensional and turbulent, which are greatly challenging for the network model. Therefore, it deserves paying more attention into the machine learning based reduced-order modeling and make it be able to better serve the physics or engineering community.

## 5. Conclusion

In this paper, based on the deep convolutional neural networks (CNNs), we proposed a data-driven model for predicting steady-state heat convection of a hot object with arbitrary complex geometry in a two-dimensional space, and the signed distance function (SDF) is applied to represent the geometry of the problem. The dataset using the hot objects with simple geometries is used to train and validation the model, including triangles, quadrilaterals, pentagons, hexagons and dodecagons. According to the results, the velocity and temperature field of the problems with complex geometries predicted by the network model agrees well with the CFD simulation. As the location and the number of the hot objects varies, the network model is still able to give a satisfied prediction. In addition, further incorporating the velocity information into the input matrix will moderately improve the performance of the network model on temperature field prediction. Although the training process takes about 42 minutes in current work, the time cost of the CNNs to predict the heat convection fields is negligible, specifically the prediction speed of the network model is four orders faster. Therefore, it is potentially possible that using this approach, designers can generate numerous design alternatives in a negligible time during the early design stage, or this approach can provide some enlightenment for real-time simulation during some control tasks.

**Acknowledgement**

This work is supported by Natural Science Foundation of China No. 11802135, the Fundamental Research Funds for the Central Universities No. 30919011401.